\documentclass[aps,twocolumn,10pt,superscriptaddress,prb]{revtex4-1}

\usepackage{amsmath,amssymb,amsfonts}
\usepackage{graphicx}
\usepackage{braket}
\usepackage{nicefrac}
\usepackage{multirow}
\usepackage[thinlines]{easytable}
\usepackage{cellspace}
\usepackage[colorlinks=True,linkcolor=red,citecolor=blue,urlcolor=blue]{hyperref}
\usepackage[retainorgcmds]{IEEEtrantools}
\usepackage{xcolor}
\usepackage[title]{appendix}
\usepackage{appendix}

\begin{document}


\title{Computing observables without eigenstates: applications to Bloch Hamiltonians}

\author{Oscar Pozo}
\affiliation{Instituto de Ciencia de Materiales de Madrid,
and CSIC, Cantoblanco, 28049 Madrid, Spain}
\author{Fernando de Juan}
\affiliation{Donostia International Physics Center, 20018 Donostia-San Sebastian, Spain}
\affiliation{IKERBASQUE, Basque Foundation for Science, Maria Diaz de Haro 3, 48013, Bilbao, Spain}

\begin{abstract}

Calculating the observables of a Hamiltonian requires taking matrix elements of operators in the eigenstate basis. Since eigenstates are only defined up to arbitrary phases that depend on Hamiltonian parameters, analytical expressions for observables are often difficult to simplify. In this work, we show how for small Hilbert space dimension $N$ all observables can be expressed in terms of the Hamiltonian and its eigenvalues using the properties of the SU($N$) algebra, 
and we derive explicit expressions for $N=2,3,4$. Then we present multiple applications specializing to the case of Bloch electrons in crystals, including the computation of Berry curvature, quantum metric and orbital moment, as well as a more complex observable in non-linear response, the linear photogalvanic effect (LPGE). As a physical example we consider multiband Hamiltonians with nodal degeneracies to show first how constraints between these observables are relaxed when going from two to three-band models, and second how quadratic dispersion can lead to constant LPGE at small frequencies.  \end{abstract}

\maketitle

\section{Introduction}


In the theory of quantum mechanics, there is a well-defined prescription to compute the physical properties of a quantum system that can be experimentally measured, known as its observables. Such system is governed by a Hamiltonian $H$, and its observables are constructed as matrix elements of operators in the basis of eigenstates of $H$. The eigenstates are only defined up to a gauge transformation that changes their phase, and therefore they are not observable. 
On the other hand, observables are by construction gauge invariant quantities. 





In practice, the analytical computation of observables through eigenstates might entail unnecessary complication because of this gauge freedom. When the Hamiltonian under consideration depends on many parameters, judicious gauge choices for the eigenstates might be required, especially if the observables require taking derivatives with respect to those parameters, or if the eigenstates become singular at degeneracy points. Numerical calculations often face a related problem as the numerical derivative is not well-defined if the phase of nearby states is arbitrary. Observables are always well-defined regardless of the method of computation, but a method that does not require the direct evaluation of gauge dependent eigenstates would be advantageous in simplifying certain calculations.


The aim of this work is to show how observables can be computed without explicit reference to eigenstates. For a general Hamiltonian of dimension $N$, we show it is possible to compute a minimal set of matrix elements that exhaust all the possible observables by using the properties of the SU($N$) algebra. 




While our method is general, the main focus of our work will be its application to Bloch Hamiltonians for electrons in a periodic potential, where $H$ depends on the crystal momentum $\vec k$. Effective Bloch Hamiltonians of small size can often be used to describe many systems of interest, and the $N=2$ is simple enough to afford analytical solutions. However, the simplicity of two-band Hamiltonians also implies that there are constraints between observables that should be in principle independent. Examples of these constraints include the fact that the Berry curvature $\vec \Omega$ and the orbital magnetic moment $\vec m$ satisfy $\vec \Omega \times \vec m=0$, or that in 2D the determinant of the quantum metric $g_{ij}$ satisfies $\det g = \Omega^2$. Another interesting example comes from non-linear optics, where there is a relation between the
shift current, a contribution to the lineal photogalvanic effect (LPGE), and the derivatives of the Berry curvature. Using our general solutions, in this work we present the calculation of all of these observables for specific Hamiltonians of dimension $N=3$, where these constraints are explicitly relaxed.


A concrete physical motivation of our work is the modeling of a particular example of Bloch Hamiltonians featuring nodal band degeneracies, for which our method is particularly useful. The simplest of these is the two-band point-like touching, known as a Weyl point~\cite{Armitage18}, but more recently all multiband touching points protected by symmetry have been classified~\cite{BradlynEA17,TangEA17,ChangEA17,Flicker18} and a certain class of them has been found experimentally in several compounds~\cite{Takane19,Sanchez19,Rao19,Schroeter19,schroeter20}. As explained in Ref.~\onlinecite{Flicker18}, all multifold band crossings realized in chiral crystals can be described at low energies by linear Hamiltonians of size $N=2,3,4$, so our results can be used rather naturally to make predictions for this class of systems. 

Our work is organized as follows. First, to motivate the logic of our method, we present a concise summary of the $N=2$ case with minimal formalism in Section~\ref{sec:StatementProblem}. Next, in Section~\ref{sec:GeneralStrategy} we discuss how such procedure can be extended to higher dimensional Hamiltonians. In Section~\ref{sec:SU(3)} we derive the general results for $N=3$. 
The results for $N=4$ follow similarly with more involved expressions that we detail in Appendix~\ref{sec:SU(4)}. 
Finally, in Section~\ref{sec:PhysicalApplications} we consider all the different physical applications described in the introduction for specific Hamiltonians with nodal degeneracies, and in Section~\ref{sec:Discussion} we discuss more general implications of our work and present our conclusions.

\section{Statement of the problem}
\label{sec:StatementProblem}
Consider a general hermitian Hamiltonian $\mathcal{H}$ of dimension $N$, and define the generators of SU($N$), $M_{\alpha}$, with $\alpha=\left\{ 1,2,...,N^{2}-1 \right\}$, normalized as $\mathrm{Tr}\left[ M_{\alpha}M_{\beta} \right] = 2\delta_{\alpha\beta}$. The Hamiltonian is expressed as
\begin{equation}
    \mathcal{H} = h_0 \mathbb{I}+ h_{\alpha}M_{\alpha}  \  ,
    \label{eq:HamNotation}
\end{equation}
where $\mathbb{I}$ is the identity matrix, $h_0,h_{\alpha}$ are functions of the parameters of the Hamiltonian and repeated SU($N$) indices $\alpha,\beta,\dots$ are always summed over. Since $h_0\mathbb{I}$ is a trivial energy shift, it is useful to separate $\mathcal{H} = h_0\mathbb{I} + H$ with
\begin{align}
    H= h_{\alpha}M_{\alpha}  \ ,
    \label{eq:HamNotation}
\end{align}
which has the eigenvalue equation
\begin{align}
 H\ket{n} = \varepsilon_{n}\ket{n}  \ , 
    \label{eq:HamEigen}
\end{align}
where $n=\left\{ 1,2,...,N \right\}$ labels the eigenstates $\ket{n}$ and eigenvalues $\varepsilon_n$. The eigenstates $\ket{n}$ are also eigenstates of $\mathcal{H}$ with eigenvalues $E_n = h_0 + \varepsilon_{n}$. We assume that all $\varepsilon_n$ are non-degenerate. 


Any traceless operator $O$ in the Hilbert space of $\mathcal{H}$ can be expressed in terms of the SU($N$) generators as $O = O_\alpha M_\alpha$, with matrix elements in the eigenstate basis
\begin{equation}
    O^{nm} \equiv \bra{n}O\ket{m}  = O_\alpha M^{nm}_\alpha \ ,
    \label{eq:MatrixElementNotation}
\end{equation}
while an identity operator has trivial matrix elements. Note $M_{\alpha}^{nm}$ itself is not invariant under a gauge transformation of the form
\begin{equation}
    \ket{n} \rightarrow e^{i\theta}\ket{n} \ ,
\end{equation}
except for $n=m$. All gauge invariant observables can be built from combinations of $M_{\alpha}^{nm}$ as strings $M_\alpha^{nm} M_\beta^{ml}\cdots M_\gamma^{pn}$ where every index $n,m,...$ appears 
an equal number of times
on the left and on the right, the simplest of these being the diagonal $M_{\alpha}^{nn}$. We will define the $R$-generators as the irreducible gauge invariant products of $R$ matrix elements $M_{\alpha}^{nm}$ that cannot be decomposed into smaller gauge invariant products. For a given $N$ there are $R$-generators with $R=\left\{ 1,\dots,N \right\}$, and they are listed explicitly in Table \ref{tab:R-generators} up to $N=4$. 



The aim of this work is to show how all $R$-generators, and hence all observables, can be computed without diagonalizing $H$. The rationale behind this is best illustrated with the simplest case, $N=2$, where the $R$-generators are $M_{\alpha}^{nn}$ and $M_{\alpha}^{12}M_{\beta}^{21}$. The defining equation of the SU(2) group algebra
\begin{align}
    M_{\alpha}M_{\beta} =  \delta_{\alpha\beta}\mathbb{I} + i\epsilon_{\alpha\beta\gamma} M_{\gamma}  \  ,   \label{eq:ProductIdentity}
\end{align}
combined with Eqs.~\eqref{eq:HamNotation} and \eqref{eq:HamEigen} provide explicit expressions for these minimal matrix elements as a function of $h_{\alpha}$ and $\varepsilon_{n}$. Here $\epsilon_{\alpha\beta\gamma}$ is the Levi-Civita symbol.
First, the 1-generators $M_{\alpha}^{nn}$ are
obtained by contracting the diagonal matrix element of Eq.~\eqref{eq:ProductIdentity} with $h_\beta$
\begin{align}
h_\beta \bra{n} M_{\alpha}M_{\beta} \ket{n} =  \varepsilon_n  M_{\alpha}^{nn} = h_{\alpha} + i \epsilon_{\alpha\beta\gamma}h_{\beta}M_{\gamma}^{nn} \ . \label{eq:MainEqSU2}
\end{align}
Since $M_{\alpha}^{nn}$ is real, Eq.~\eqref{eq:MainEqSU2} leads to 
\begin{equation}
    M_{\alpha}^{nn} = \dfrac{h_{\alpha}}{\varepsilon_{n}}  \ .\label{eq:SU2_1}
\end{equation}
Second, the 2-generator $M_{\alpha}^{12}M_{\beta}^{21}$ 
is obtained by introducing a resolution of the identity $\mathbb{I} = \sum_m\ket{m} \bra{m}$ in the diagonal matrix element of Eq. \eqref{eq:ProductIdentity}, giving
\begin{IEEEeqnarray}{rCl}
     M_{\alpha}^{12}M_{\beta}^{21}  & = & \delta_{\alpha\beta} - M_\alpha^{11}M_\beta^{11} +i \epsilon_{\alpha\beta\gamma}M_{\gamma}^{11} \ . \label{eq:SU2_2}
\end{IEEEeqnarray}
Finally, the condition $\det(H-\varepsilon_{n}\mathbb{I})=0$ determines the analytic expression of the energies
\begin{equation} 
    \varepsilon_{n} = \left( -1 \right)^{n} \sqrt{h_\alpha h_\alpha}  \  . \label{eq:SU2_En}
\end{equation}
Equations \eqref{eq:SU2_1}-\eqref{eq:SU2_En} solve the SU(2) problem completely: any possible observable can be built from them, and diagonalizing $H$ or explicit eigenstates were never required to find them. 




\section{General strategy}
\label{sec:GeneralStrategy}


We now consider problem for SU($N$), and explain the strategy to obtain the $R$-generators using SU($N$) identities. The Lie algebra of SU($N$) is characterized by the completely symmetric (antisymmetric) structure constants $d_{\alpha\beta\gamma}$ ($f_{\alpha\beta\gamma}$) defined as
\begin{IEEEeqnarray}{rCl}
    d_{\alpha\beta\gamma} & = & \dfrac{1}{4} \text{Tr}\Big[ M_{\alpha}\left\{ M_{\beta}, M_{\gamma} \right\} \Big]  \  ,  \label{eq:dDef} \\
    f_{\alpha\beta\gamma} & = & -\dfrac{i}{4} \text{Tr}\Big[ M_{\alpha}\left[ M_{\beta}, M_{\gamma} \right] \Big]  \  , \label{eq:fDef}
\end{IEEEeqnarray}
which usually appear in the form
\begin{equation}
    S_{\alpha\beta\gamma}\equiv if_{\alpha\beta\gamma}+d_{\alpha\beta\gamma}  \  .
\end{equation}
These structure constants obey the following identities~\cite{Macfarlane}
\begin{IEEEeqnarray}{rCl}
    0 & = & f_{\alpha\sigma\rho}d_{\rho\beta\gamma} + f_{\beta \sigma\rho}d_{\alpha\rho\gamma} + f_{\gamma \sigma\rho}d_{\alpha\beta\rho}  \  ,  \label{eq:fdCyclic}  \\
    0 & = & f_{\alpha\sigma\rho}f_{\rho\beta\gamma} + f_{\beta \sigma\rho}f_{\alpha\rho\gamma} + f_{\gamma \sigma\rho}f_{\alpha\beta\rho}  \  , \label{eq:ffCyclic}  \\
    f_{\alpha\beta\sigma}f_{\gamma\delta\sigma} & = & \dfrac{2}{N}\left( \delta_{\alpha\gamma}\delta_{\beta\delta}-\delta_{\alpha\delta}\delta_{\beta\gamma} \right) + \nonumber \\ 
    & & + d_{\alpha\gamma\sigma}d_{\beta\delta\sigma}-d_{\beta\gamma\sigma}d_{\alpha\delta\sigma}  \  . \label{eq:ff-dd}
\end{IEEEeqnarray}
In addition, the SU($N$) algebra has a defining equation which is the central object to determine the $R$-generators
\begin{equation}
     M_{\alpha}M_{\beta} = \dfrac{2}{N} \delta_{\alpha\beta}\mathbb{I} + S_{\alpha\beta\gamma} M_{\gamma}  \  . \label{eq:MainEq} 
\end{equation}
In the $N=2$ case discussed above we simply have $d_{\alpha \beta \gamma} =0$, while $f_{\alpha \beta \gamma} =\epsilon_{\alpha \beta \gamma}$, recovering the familiar algebra \eqref{eq:ProductIdentity} of the Pauli matrices.

We introduce for convenience a family of tensors $d_{\alpha_{1}\cdots\alpha_{r}}^{(r)}$ defined through recursive contractions of the symmetric structure constants
\begin{IEEEeqnarray}{rCl}
    d^{(2)}_{\alpha_{1}\alpha_{2}} & = & \delta_{\alpha_{1}\alpha_{2}}  \ ,  \\
    d^{(3)}_{\alpha_{1}\alpha_{2}\alpha_{3}} & = & d_{\alpha_{1}\alpha_{2}\alpha_{3}}  \  ,  \\
    d^{(r)}_{\alpha_{1}\cdots \alpha_{r}} & = & d^{(r-1)}_{\alpha_{1}\cdots \alpha_{r-2}\sigma}d_{\sigma \alpha_{r-1}\alpha_{r}}  \  ,  \quad 4\leq r \leq N  \  ,
\end{IEEEeqnarray}
which is a generalization of the completely symmetric $d$-family~\cite{de1998invariant}. These tensors can be used to construct the following SU($N$) scalars
\begin{IEEEeqnarray}{rCl}
    s_{r} & \equiv & d^{(r)}_{\alpha_{1}\cdots \alpha_{r}}h_{\alpha_{1}}\cdots h_{\alpha_{r}}  \ ,  \quad 2\leq r \leq N  \  , \label{eq:SUN_Scalars}
\end{IEEEeqnarray}
and SU($N$) vectors
\begin{IEEEeqnarray}{rCl}
    V_{\alpha}^{(r)} & \equiv & d^{(r+1)}_{\alpha\beta_{1}\cdots \beta_{r}}h_{\beta_{1}}\cdots h_{\beta_{r}}  \ ,  \quad 1\leq r \leq N-1  \  . \label{eq:SUN_Vectors}
\end{IEEEeqnarray}
In addition to Eqs. \eqref{eq:fdCyclic}-\eqref{eq:ff-dd}, for each value of $N$ there is a specific identity involving the $d$-family tensors~\cite{de1998invariant,rashid1973identity}. We will refer to this identity as the closing $d$-family identity, 
because it closes the recursive definitions \eqref{eq:SUN_Scalars}-\eqref{eq:SUN_Vectors}, since it implies that any scalar $s_{r}$ with $r>N$ and any vector $V_{\alpha}^{(r)}$ with $r\geq N$ can be written in terms of the previous ones. The specific identities for $N=3,4$ are shown in Sec.~\ref{sec:SU(3)} and in the Appendix \ref{sec:SU(4)}, respectively.

\begin{figure}[!]
    \centering
    \includegraphics{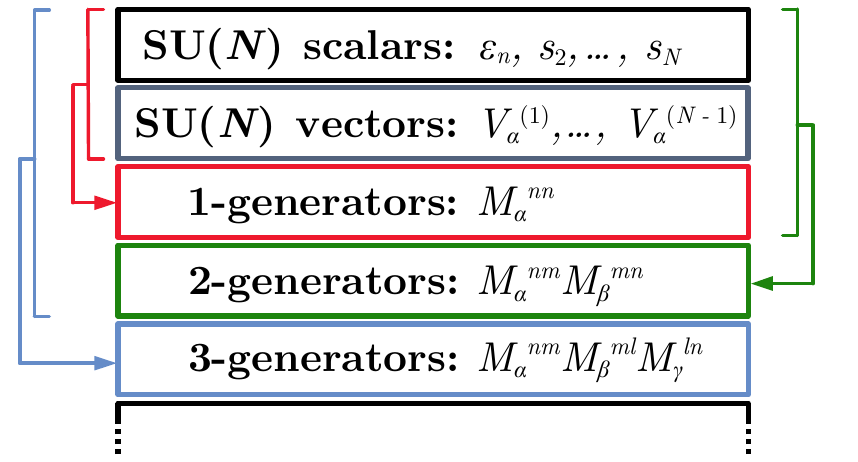}
    \caption{Schematic illustration of the general guideline showing the elements required to compute the $R$-generators. $(n,m,l)$ indices are assumed to be different.}
    \label{fig:GuidelineScheme}
\end{figure}

\subsection{Energies}
\label{sec:Energies}
The Hamiltonian energies are essential to determine the $R$-generators, but these matrix elements do not explicitly depend on $h_0$. This is general, and follows from the fact that the states $\ket{n}$ are independent of $h_{0}$. Thus we only need $H$, and not $\mathcal{H}$ in what follows.
The energies $\varepsilon_{n}$ are completely determined by the roots of the characteristic polynomial for $H$~\cite{roman2005advanced}

\begin{equation}
    \det\left( H - \varepsilon_{n}\mathbb{I} \right) \equiv p_{H}(N,\varepsilon_{n}) = 0 \ .
    \label{eq:CharacPol}
\end{equation}
This polynomial can be obtained explicitly by expressing the determinant in a sum of powers of ${\rm tr}(H- \varepsilon_{n}\mathbb{I})^n$ with $n=1,\dots,N$~\cite{mehta2004random}, which can then be rearranged in terms of the scalars $s_n$. Explicit examples of this are shown in Secs.~\ref{sec:SU(3)} and \ref{sec:SU(4)} respectively, where the roots of these polynomials can be obtained analytically. For larger values of $N$ there is no general analytical solution, but some simplifications might occur for certain choices of $h_\alpha$. 
\begin{table*}[t]
\setlength{\tabcolsep}{1.75pt} 
\renewcommand{\arraystretch}{1.5} 
\begin{center}
    \begin{tabular}{ c | c c c c } 
        \hline
        \ \textbf{\textit{N}} \ & \ \ \textbf{1-generators} \ \ & \ \ \textbf{2-generators} \ \ & \ \ \textbf{3-generators} \ \ & \ \ \textbf{4-generators} \ \ \\ [0pt]
        \hline
        \hline
        2 & $M_{\alpha}^{11}$, $M_{\alpha}^{22}$ & $M_{\alpha}^{12}M_{\beta}^{21}$ & - & - \\ [1pt]
        \hline
        3 & \ \ $M_{\alpha}^{11}$, $M_{\alpha}^{22}$, $M_{\alpha}^{33}$ \ & \ $M_{\alpha}^{12}M_{\beta}^{21}$, $M_{\alpha}^{23}M_{\beta}^{32}$, $M_{\alpha}^{31}M_{\beta}^{13}$ \ & \ $M_{\alpha}^{12}M_{\beta}^{23}M_{\gamma}^{31}$ \ & - \\ [1pt] 
        \hline
        \multirow{2}{4pt} 4 & $M_{\alpha}^{11}, M_{\alpha}^{22},$ & $M_{\alpha}^{12}M_{\beta}^{21}$, $M_{\alpha}^{13}M_{\beta}^{31}$, $M_{\alpha}^{14}M_{\beta}^{41}$, \ & \ $M_{\alpha}^{12}M_{\beta}^{23}M_{\gamma}^{31}$, $M_{\alpha}^{12}M_{\beta}^{24}M_{\gamma}^{41}$, \ & \ $M_{\alpha}^{12}M_{\beta}^{23}M_{\gamma}^{34}M_{\delta}^{41}$, $M_{\alpha}^{12}M_{\beta}^{24}M_{\gamma}^{43}M_{\delta}^{31}$, \ \\ [-1pt]
        & \ $M_{\alpha}^{33}$, $M_{\alpha}^{44}$ \ & \ $M_{\alpha}^{23}M_{\beta}^{32}$, $M_{\alpha}^{24}M_{\beta}^{42}$, $M_{\alpha}^{34}M_{\beta}^{43}$ \ & \ $M_{\alpha}^{13}M_{\beta}^{34}M_{\gamma}^{41}$, $M_{\alpha}^{23}M_{\beta}^{34}M_{\gamma}^{41}$ \ & \ $M_{\alpha}^{13}M_{\beta}^{32}M_{\gamma}^{24}M_{\delta}^{41}$ \ \\ [1pt]
        \hline
    \end{tabular}
    \caption{Set of $R$-generators for SU(2), SU(3) and SU(4).}
    \label{tab:R-generators}
\end{center}
\end{table*}

\subsection{1-generators}
\label{sec:1-gen}


The 1-generators $M_{\alpha}^{nn}$ are diagonal and real by construction. To determine their analytic expression we evaluate the real part of $\bra{n}M_{\alpha}H\ket{n}$ using Eq.~\eqref{eq:HamEigen} and Eq.~\eqref{eq:MainEq} separately. This leads to the relation
\begin{equation}
    \varepsilon_{n}M_{\alpha}^{nn} = \dfrac{2}{N}h_{\alpha} + d_{\alpha\beta\gamma}h_{\beta}M_{\gamma}^{nn}  \ .
    \label{eq:dhM}
\end{equation}
Since $M_{\alpha}^{nn}$ is an SU($N$) vector, and the only vectors one can build with $h_\alpha$ are those given in Eq.~\eqref{eq:SUN_Vectors}, we decompose $M_{\alpha}^{nn}$ as
\begin{equation}
    M_{\alpha}^{nn} = \sum_{r=1}^{N-1} X_{n}^{(r)} V_{\alpha}^{(r)} \ ,
    \label{eq:VectorDecomposition}    
\end{equation}
and introduce the vector decomposition in Eq.~\eqref{eq:dhM} to find the coefficients $X_{n}^{(r)}$ and then the 1-generators. 
This is achieved by using the tensor contractions
\begin{equation}
    d_{\alpha\beta\gamma}h_{\beta}V_{\gamma}^{(r)} = V_{\alpha}^{(r+1)} \ , \qquad r = 1,\dots, N-2 \ ,
\end{equation}
which follow from Eq.~\eqref{eq:SUN_Vectors}, and the contraction $d_{\alpha\beta\gamma}h_{\beta}V_{\gamma}^{(N-1)}$, which requires the closing $d$-family identity for the specific value of $N$.

 
It should be noted that instead of contracting $\bra{n}M_{\alpha}M_{\beta}\ket{n}$ with $h_{\beta}$ to get Eq.~\eqref{eq:dhM} we could have contracted with any other vector $V_{\alpha}^{(r)}$ to get a different relation. All these equations are equivalent since each of them unequivocally determines $M_{\alpha}^{nn}$. 



\subsection{2-generators}
\label{sec:2-gen}






There are many possible tensors that can be constructed from $h_{\alpha}$ and the structure constants~\cite{Macfarlane}, so 
a tensor decomposition of $R$-generators analogous to \eqref{eq:VectorDecomposition} becomes too impractical for $R\geq2$.
A better option relies on 
the diagonal elements of Eq.~\eqref{eq:MainEq}, which relate 2-generators with 1-generators as follows
\begin{equation}
    \sum_{m=1}^{N}M_{\alpha}^{nm}M_{\beta}^{mn} = \dfrac{2}{N}\delta_{\alpha\beta} + S_{\alpha\beta\gamma}M_{\gamma}^{nn} \ .
    \label{eq:Easiest-2gen-1gen}
\end{equation}
Similarly, larger products of $M_{\alpha}$ matrices lead to different relations between 2-generators and 1-generators if the remaining SU($N$) indices are contracted with the vectors \eqref{eq:SUN_Vectors}. There are as many relations as different 2-generators, so it is always possible to find a set of equations that completely determines them. However, these sets can only determine the real part $\mathrm{Re}\big[ M_{\alpha}^{nm}M_{\beta}^{mn} \big]$. This is because the imaginary part of Eq.~\eqref{eq:Easiest-2gen-1gen} produces a homogeneous system of equations (with no constant term) for $\mathrm{Im}\big[ M_{\alpha}^{nm}M_{\beta}^{mn} \big]$. Nevertheless, these imaginary parts can be determined using the non-diagonal matrix element $\bra{n} \big[ M_{\alpha}, H \big] \ket{m}$, which gives
\begin{IEEEeqnarray}{rCl}
    M_{\alpha}^{nm} & = & \dfrac{2if_{\alpha\sigma\rho}}{\varepsilon_{m}-\varepsilon_{n}} h_{\sigma}M_{\rho}^{nm}  \ .
    \label{eq:GaugeDepReIm}
\end{IEEEeqnarray}
Multiplying this gauge dependent equation by $M_{\beta}^{mn}$, we find the imaginary part of 2-generators in terms of the real ones
\begin{equation}
    \mathrm{Im}\big[ M_{\alpha}^{nm}M_{\beta}^{mn} \big] = \dfrac{2f_{\alpha\sigma\rho}h_{\sigma}}{\varepsilon_{m}-\varepsilon_{n}} \mathrm{Re} \big[ M_{\rho}^{nm}M_{\beta}^{mn} \big]  \  .
    \label{eq:ReMM-to-ImMM}
\end{equation}

\subsection{Higher order $R$-generators}
\label{sec:R-gen}


It is possible to obtain $R$-generators with $R\geq3$ following the logic of the previous section but considering larger products of SU($N$) generators together with Eq.~\eqref{eq:MainEq}. For example, the product $M_{\alpha}M_{\beta}M_{\gamma}$ provides the identity
\begin{IEEEeqnarray}{rCl}
    \sum_{m,l=1}^{N} M_{\alpha}^{nm}M_{\beta}^{ml}M_{\gamma}^{ln} & = & \dfrac{2}{N}\left( \delta_{\alpha\beta}M_{\gamma}^{nn} + S_{\alpha\beta\gamma} \right) + \nonumber \\
    & + & S_{\alpha\beta\sigma}S_{\sigma\gamma\rho}M_{\rho}^{nn} \ ,
\end{IEEEeqnarray}
we obtain relations between 3- and 1-generators, but they become too impractical to solve.
For $R\geq3$ there is an alternative that is not present when considering 2-generators. It consists in taking non-diagonal expectation values of Eq.~\eqref{eq:MainEq} instead of diagonal ones to reach gauge dependent relations as
\begin{equation}
    \sum_{m=1}^{N} M_{\alpha}^{nm}M_{\beta}^{ml} = S_{\alpha\beta\sigma}M_{\sigma}^{nl} \ ,
\end{equation}
where $l\neq n$. This equation may be made gauge invariant multiplying it by any string of matrix elements that compensates the gauge dependence, for instance $M_{\gamma}^{ln}$, which provides a relation between 2- and 3-generators
\begin{equation}
    \sum_{m=1}^{N} M_{\alpha}^{nm}M_{\beta}^{ml}M_{\gamma}^{ln} = S_{\alpha\beta\sigma}M_{\sigma}^{nl}M_{\gamma}^{ln} \ .
    \label{eq:Easiest-3gen-2gen}
\end{equation}
As for 2-generators, different relations arise for $R\geq3$ when considering larger products in which the remaining SU($N$) indices are contracted with the vectors \eqref{eq:SUN_Vectors}. Their real and imaginary parts are related by Eq.~\eqref{eq:GaugeDepReIm} when it is multiplied by a suitable gauge dependent string, for instance $M_{\beta}^{ml}M_{\gamma}^{ln}$
\begin{equation}
    \mathrm{Im}\big[ M_{\alpha}^{nm}M_{\beta}^{ml}M_{\gamma}^{ln} \big] = \dfrac{2if_{\alpha\sigma\rho}h_{\sigma}}{\varepsilon_{m}-\varepsilon_{n}} \mathrm{Re}\big[ M_{\rho}^{nm}M_{\beta}^{ml}M_{\gamma}^{ln} \big] . \quad
    \label{eq:ReMMM-to-ImMMM}
\end{equation}

As a final practical comment on the general procedure, it is worth noting that the simplest expressions for $R$-generators for $N>2$ are those expressed in terms of lower order generators, as opposed to explicit expressions in terms $h_\alpha$, $\varepsilon_n$ and the structure constants, or in terms of the scalars \eqref{eq:SUN_Scalars} and vectors \eqref{eq:SUN_Vectors}. Because of this we have not presented such type of expressions in the text.

\section{SU(3) solution}
\label{sec:SU(3)}
In this section we apply the general strategy to the case $N=3$. The starting point is the closing $d$-family identity for $N=3$, which is~\cite{de1998invariant}
\begin{IEEEeqnarray}{rCl}
    d^{(4)}_{\left( \alpha\beta\gamma\delta \right)} & = & \dfrac{1}{3}\delta_{(\alpha\beta}\delta_{\gamma\delta)}  \ , \label{eq:SU(3)-d-Identity} \\
    d_{(\alpha_{1}\cdots\alpha_{r})}^{(r)} & \equiv & \dfrac{1}{N_{\mathcal{P}}}\sum_{\mathcal{P}}d_{\mathcal{P}\alpha_{1}\cdots\mathcal{P}\alpha_{r}}^{(r)} \ ,
    \label{eq:SymDef}
\end{IEEEeqnarray}
where $\sum_{\mathcal{P}}$ denotes the sum over the permutations of $\left\{ \alpha_{1},\dots,\alpha_{r} \right\}$ and $N_{\mathcal{P}}$ is the number of different permutations. The characteristic polynomial \eqref{eq:CharacPol} for $N=3$,
\begin{equation}
    p_{H}(3,\varepsilon_{n}) = -\varepsilon_{n}^{3} + s_{2}\varepsilon_{n} + \dfrac{2}{3}s_{3} \ , \label{eq:ChaEqSU3}
\end{equation}
leads to the following analytic expression for the energies~\cite{nickalls2006viete}
\begin{equation}
    \varepsilon_{n} = 2\sqrt{\dfrac{s_{2}}{3}} \cos \left[ \dfrac{1}{3}\arccos\left( \dfrac{s_{3}}{s_{2}}\sqrt{\dfrac{3}{s_{2}}} \right) + \dfrac{2\pi n}{3} \right] \ .
\end{equation}

To determine the 1-generators, there are only two vectors available. We need the specific tensor contraction
\begin{IEEEeqnarray}{rCl}
    d_{\alpha\beta\gamma}h_{\beta}V_{\gamma}^{(2)} & = & \dfrac{s_{2}}{3}V_{\alpha}^{(1)} \ ,
\end{IEEEeqnarray}
which comes from the $d$-family identity \eqref{eq:SU(3)-d-Identity}. Then the vector decomposition of $M_{\alpha}^{nn}$ in Eq.~\eqref{eq:dhM} leads to the following expression for the 1-generators
\begin{equation}
    M_{\alpha}^{nn} = 2\dfrac{\varepsilon_{n}V_{\alpha}^{(1)}+V_{\alpha}^{(2)}}{3\varepsilon_{n}^{2}-s_{2}}  \ .
    \label{eq:SU(3)-M}
\end{equation}
It is worth mentioning that the contraction $V_{\beta}^{(2)}\bra{n}M_{\alpha}M_{\beta}\ket{n}$ gives directly the form of $M_{\alpha}^{nn}$ without using tensor decomposition.

The real part of the 2-generators is fully determined by the real part of Eq.~\eqref{eq:Easiest-2gen-1gen}. There are three possible 2-generators and three equations. The solution is
\begin{IEEEeqnarray}{rCl}
    \mathrm{Re}\left[ M_{\alpha}^{nm}M_{\beta}^{mn} \right] & = & \dfrac{1}{3}\delta_{\alpha\beta} +  d_{\alpha\beta\gamma}\left( M_{\gamma}^{nn}+M_{\gamma}^{mm} \right) + \nonumber \\
    & & + \dfrac{1}{2}\left( M_{\alpha}^{nn}M_{\beta}^{mm}+M_{\alpha}^{mm}M_{\beta}^{nn} \right) .  \qquad
\end{IEEEeqnarray}
The imaginary part of the 2-generators is obtained using Eq.~\eqref{eq:ReMM-to-ImMM}.

Finally, there is only one 3-generator whose real part can be determined from the real part of Eq.~\eqref{eq:Easiest-3gen-2gen} choosing $(n,m)=(1,3)$. It reads
\begin{IEEEeqnarray}{rCl}
\mathrm{Re} & & \left[ M_{\alpha}^{12}M_{\beta}^{23}M_{\gamma}^{31} \right] = \nonumber\\ 
& & d_{\alpha\beta\sigma}\mathrm{Re}\left[ M_{\sigma}^{31}M_{\gamma}^{13} \right] +f_{\alpha\beta\sigma}\mathrm{Im}\left[ M_{\sigma}^{31}M_{\gamma}^{13} \right] - \nonumber \\    
& & - M_{\alpha}^{11}\mathrm{Re}\left[ M_{\beta}^{31}M_{\gamma}^{13} \right] - M_{\gamma}^{33}\mathrm{Re}\left[ M_{\alpha}^{31}M_{\beta}^{13} \right]  ,
\end{IEEEeqnarray}
while the imaginary part is given by Eq.~\eqref{eq:ReMMM-to-ImMMM}. 

\section{Application to Bloch Hamiltonians}
\label{sec:PhysicalApplications}

For the rest of this work, we consider Bloch Hamiltonians for electrons moving in a crystal with a periodic potential~\cite{ashcroft1976solid}. The dimension $N$ in this case corresponds to the number of orbitals in the unit cell, or the relevant low-energy states in a $\vec{k}\cdot \vec{p}$ approximation, and the Hamiltonian is a function of the crystal momentum, $H_{0}(\vec k)$. We use latin indices to indicate cartesian components such as $k_a$, with $a=\left\{ 1,2,3 \right\}$. 

Specifically, we will focus on computing a series of observables for Hamiltonians with nodal degeneracies, where the wavefunctions become singular at a single point in momentum space. Our method applies to observables that are evaluated at finite momenta with respect to the node, so that $\varepsilon_n$ is always non-degenerate. Our method avoids computing such wavefuncions or their derivatives, and produces the observables of interest directly from the 
$R$-generators, which are not singular. The simplest of such Hamiltonians is a Weyl point, a linear touching of two bands~\cite{Armitage18}, for which many observables can be computed analytically thanks to the simplicity of Eqs. \eqref{eq:SU2_1}-\eqref{eq:SU2_En}. 
For illustrative purposes
we will rather consider the novel features that arise with three-band Hamiltonians. 

\begin{figure}[!t]
    \centering
    \includegraphics*[scale=1.0]{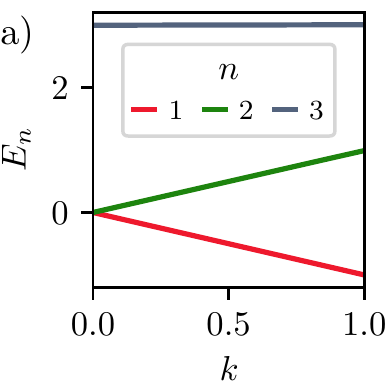}
    \includegraphics*[scale=1.0]{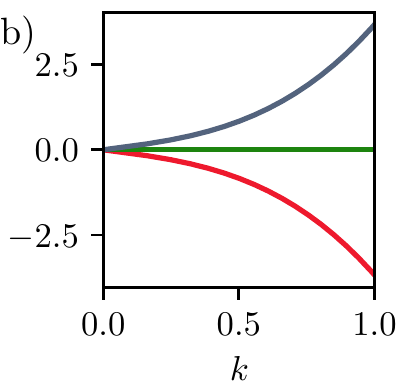}
    \caption{a) Energy bands of the Weyl fermion with an extra band, Eq.~\eqref{eq:Ham1}, with $(a,c,\Delta)=(1.0,0.1,3.0)$. b) Energy bands of the threefold fermion with quadratic corrections from the point group $T$, Eq.~\eqref{eq:Ham2}, with $(a,d)=(1.0,2.0)$. In both examples we show the direction $\vec{k}=(k,0,0)$.}
    \label{fig:EnergyBands}
\end{figure}

The first Hamiltonian we consider is that of a Weyl fermion with an extra band located at an energy $\Delta$ above the Weyl node
\begin{equation}
    H_{W} = \left( \begin{matrix} a k_{z} - \dfrac{\Delta}{3} & b(k_{x}+ik_{y}) & c(k_{x}-ik_{y}) \\ b(k_{x}-ik_{y}) & -a k_{z} - \dfrac{\Delta}{3} & c(k_{x}+ik_{y}) \\ c(k_{x}+ik_{y}) & c(k_{x}-ik_{y}) & \dfrac{2\Delta}{3} \end{matrix} \right) \ ,
    \label{eq:Ham1}
\end{equation}
in which the three bands present $C_{3}$ symmetry and explicitely break time-reversal symmetry. For simplicity, we will consider the case $b=a$, and the chemical potential will always lie at the node. The coupling to the extra band can induce new contributions to observables that are not present in the two-band model. 

The second model is the simplest multiband nodal Hamiltonian, a time-reversal symmetric threefold fermion as realized at the $\Gamma$ point in CoSi and related materials in space group 198~\cite{BradlynEA17,TangEA17,ChangEA17,Flicker18}.  The effective Hamiltonian for this system up to second order in momentum was derived in Ref.~\onlinecite{ni2020giant} and has several terms. Here, for simplicity, we will only consider one type of quadratic correction, which is the minimal one that ensures that the Hamiltonian has the symmetries of point group $T$ (corresponding to SG 198) and not higher. The Hamiltonian is  
\begin{equation}
    H_{3f} = \left( \begin{matrix} \dfrac{2d(k_{y}^{2}-k_{x}^{2})}{\sqrt{3}} & i a k_{x} & -i a k_{y} \\ -i a k_{x} & \dfrac{2d(k_{x}^{2}-k_{z}^{2})}{\sqrt{3}} & i a k_{z} \\ i a k_{y} & -i a k_{z} & \dfrac{2d(k_{z}^{2}-k_{y}^{2})}{\sqrt{3}} \end{matrix} \right) .
    \label{eq:Ham2}
\end{equation}
We will assume a chemical potential lying infinitesimally above the node.


In the following subsection we present the computation of a set of observables, the Berry curvature, quantum metric, and orbital moment. They originate from 2-generators only and afford the extra simplification that they can be written in terms of projector operators, illustrating an alternative way to use our formalism to compute observables. In the next subsection we consider a more complicated observable, the linear photogalvanic effect, which does not admit these simplifications and furthermore requires the use of both 2- and 3-generators. 

\subsection{Berry curvature, quantum metric and orbital moment}

Our first set of examples illustrates the computations of three quantities that often appear in the response functions of Bloch electrons. 
They are conveniently defined in terms of the interband matrix elements of the position operator $\vec{r}$,
\begin{align}
r_{nm}^{a} = i\bra{n}\partial_{k_a}\ket{m} \ , 
\end{align}
since they coincide with the non-diagonal connection coefficients when $n\neq m$ and are zero otherwise~\cite{sipe2000second}. The three quantities of interest are the Berry curvature
\begin{equation}
    \Omega_{ij} = \dfrac{i}{2}\sum_{\substack{n\in\mathrm{oc} \\ m\in\mathrm{unoc}}} \left( r_{nm}^{i}r_{mn}^{j} - r_{nm}^{j}r_{mn}^{i} \right) \ ,
    \label{eq:BerryCurvatureSum}
\end{equation}
the quantum metric
\begin{equation}
    g_{ij} = \dfrac{1}{2}\sum_{\substack{n\in\mathrm{oc} \\ m\in\mathrm{unoc}}} \left( r_{nm}^{i}r_{mn}^{j} + r_{nm}^{j}r_{mn}^{i} \right) \ ,
    \label{eq:QuantumMetricSum}
\end{equation}
and the orbital magnetic moment~\cite{souza2008dichroic}
\begin{equation}
    m_{ij} = \dfrac{i}{2}\sum_{\substack{n\in\mathrm{oc} \\ m\in\mathrm{unoc}}}\left( \varepsilon_{n}-\varepsilon_{m} \right) \left( r_{nm}^{i}r_{mn}^{j} - r_{nm}^{j}r_{mn}^{i} \right) \ . \label{eq:OrbitalMomentSum}
\end{equation}
They constitute a set of physical observables that can be written in terms of the projector onto occupied states, $P=\sum_{n\in\mathrm{oc}}\ket{n}\bra{n}$, as follows
\begin{IEEEeqnarray}{rCl}
    g_{ij} & = & \dfrac{1}{2}\mathrm{Tr}\Big[ \left( \partial_{k_{i}}P \right) \left( \partial_{k_{j}}P \right) \Big] \ , \\
    \Omega_{ij} & = & \dfrac{i}{2}\mathrm{Tr}\Big[ \left( \partial_{k_{i}}P \right) \left( \mathbb{I}-2P \right) \left( \partial_{k_{j}}P \right) \Big] \ , \\
    m_{ij} & = & \dfrac{i}{2}\mathrm{Tr}\Big[ H \left[ \partial_{k_{i}}P , \partial_{k_{j}}P \right] \Big] . \quad 
\end{IEEEeqnarray}
This simplifies their computation thanks to the fact that the projector $P$ is determined by the SU($N$) Fierz completeness relations
\begin{equation}
    \delta_{il}\delta_{kj} = \dfrac{1}{N}\delta_{ij}\delta_{kl} + \dfrac{1}{2}\left( M_{\alpha} \right)_{ij} \left( M_{\alpha} \right)_{kl} \ ,
\end{equation}
which lead to the following expression in terms of 1-generators
\begin{IEEEeqnarray}{rCl}
    P & = & \sum_{n\in\mathrm{oc}}\left( \dfrac{1}{N}\mathbb{I} + \dfrac{1}{2}M_{\alpha}^{nn}M_{\alpha} \right) \ .
\end{IEEEeqnarray}
The Berry curvature and the quantum metric have very simple expressions in terms of 1-generators
\begin{IEEEeqnarray}{rCl}
    \Omega_{ij} & = & -\dfrac{1}{4}\sum_{\substack{n,m,\\p\in\mathrm{oc}}} f_{\alpha\beta\gamma}M_{\alpha}^{nn}\left( \partial_{k_{i}}M_{\beta}^{mm} \right) \left( \partial_{k_{j}}M_{\gamma}^{pp} \right) , \\
    g_{ij} & = & \dfrac{1}{4}\sum_{n,m\in\mathrm{oc}} \left( \partial_{k_{i}}M_{\alpha}^{nn} \right) \left( \partial_{k_{j}}M_{\alpha}^{mm} \right) ,
\end{IEEEeqnarray}
and the Berry curvature results are consistent with those previously reported for this case~\cite{BBG12,LPG15}.

In the case of two-band Hamiltonians, these three quantities are not independent of each other, as there are several constraints between them that ultimately derive from the fact that their definitions in Eqs.~\eqref{eq:BerryCurvatureSum}, \eqref{eq:QuantumMetricSum} and \eqref{eq:OrbitalMomentSum}  contain a single term in the sum. For example, defining the quantity
\begin{equation}
G_{ijkl}=\Omega_{ij}\Omega_{kl} - g_{ik}g_{jl} - g_{jk}g_{il}\ ,
\end{equation}
an explicit substitution of Eqs. \eqref{eq:BerryCurvatureSum} and \eqref{eq:QuantumMetricSum} for a two-band model reveals that~\cite{Palumbo18}
\begin{equation}
G_{ijkl}=0\ . \label{eq:Identity1}
\end{equation}
In two dimensions, this leads to $\Omega_{xy}^2 = g_{xx}g_{yy} - g_{xy}^2 = \det g$, and in general the case $ij=kl$ can be used to recover $\Omega_{ij}$ from $g_{ij}$ up to a sign. Another two-band model identity is that the Berry curvature repackaged as a vector as $\Omega_i = \epsilon_{ijk} \Omega_{jk}$ is parallel to the orbital moment $m_i = \epsilon_{ijk}m_{jk}$ so that
\begin{equation}
    \vec{\Omega}\times\vec{m} = 0 \ .
    \label{eq:Identity2}
\end{equation}

We can now illustrate the breakdown of the identities \eqref{eq:Identity1} and \eqref{eq:Identity2} when extra bands are present. Firstly, we consider the Weyl node model with an extra band in Eq.~\eqref{eq:Ham1}, in which we can compute these corrections perturbatively in the coupling $c$ between the Weyl bands and the extra band. We obtain
\begin{IEEEeqnarray}{rCl}
    G_{xyxy} & = & \dfrac{c^{2}\left( 5+3\cos\left( 2\theta \right) - 2\cos\left( 3\phi \right) \sin\left( \theta \right)^{3} \right)}{8k^{2}\Delta^{2}} , \qquad \\
    \vec{\Omega}\times\vec{m} & = & \dfrac{4c^{2}}{\Delta}\left\{ -\dfrac{k_{y}k_{z}}{k^{2}} , \dfrac{k_{x}k_{z}}{k^{2}} , 0 \right\} ,
\end{IEEEeqnarray}
where $G_{ijkl}$ is expressed in polar coordinates $k,\theta,\phi$. These are properties of the lowest occupied band which are not present in the two-band model, and indeed vanish if either the extra band is decoupled, $c=0$, or it is infinitely far away in energy, $\Delta\rightarrow \infty$. 

Secondly, we consider the model in Eq.~\eqref{eq:Ham2} for a threefold fermion. It turns out that the minimal model for a time-reversal symmetric threefold fermion with only linear dispersion accidentally satisfies Eq.~\eqref{eq:Identity1} because the linear model has full rotational symmetry~\cite{Palumbo18}, and Eq.~\eqref{eq:Identity2} because the Berry curvature and orbital moment must be radial in any linear model~\cite{Flicker18}. However, the conditions under which the constraints \eqref{eq:Identity1} and \eqref{eq:Identity2} are satisfied do not hold once quadratic corrections are included, and in that case we obtain
\begin{widetext}
\begin{IEEEeqnarray}{rCl}
    G_{xyxy} & = & \dfrac{d^{2}\sin(\theta)^{2}\left( 35+28\cos(2\theta)+\cos(4\theta)+8\cos(4\phi)\sin(\theta)^{4} \right)}{96a^{2}k^{2}} \ , \\
    \vec{\Omega}\times\vec{m} & = & \dfrac{8d^{2}}{3ak^{5}} \left\{ k_{y}k_{z}(k_{z}^{2}-k_{y}^{2}) , k_{x}k_{z}(k_{x}^{2}-k_{z}^{2}) , k_{x}k_{y}(k_{y}^{2}-k_{x}^{2}) \right\} \ .
\end{IEEEeqnarray}
\end{widetext}

\subsection{Photogalvanic effects in multifold fermions}

In this section, we consider the linear photogalvanic effect as a physical example in which 3-generators appear. This photocurrent is generated in a non-centrosymmetric material to second order in the electric field~\cite{sipe2000second,de2020difference}. In the length gauge, the coupling of the electric field to the Bloch Hamiltonian takes the form
\begin{equation}
    H = H_{0} - e\vec{E}\cdot\vec{r} \ ,
\end{equation}
and in perturbation theory the response coefficient can be written in terms of the intraband position matrix elements as
\begin{IEEEeqnarray}{rCl}
    \sigma_{\mathrm{shift}}^{abc}(\omega) & = & \dfrac{\pi e^{3}}{2\hbar^{2}} \sum_{\substack{n\in\mathrm{oc} \\ m\in\mathrm{unoc}}} \int\dfrac{d^{3}k}{(2\pi)^{3}} f_{nm} I_{nm}^{abc} \delta(\varepsilon_{mn}-\omega) , \qquad \label{eq:ShiftCurrent} \\
    I_{nm}^{abc} & = & R_{nm}^{abc} + R_{nm}^{acb} ,
    \label{eq:ShiftIntegrand}
\end{IEEEeqnarray}
where $\varepsilon_{nm}\equiv \varepsilon_{n}-\varepsilon_{m}$, $f_{nm}=f_{n}-f_{m}$, $f_{n}$ is the Fermi-Dirac distribution function with energy $\varepsilon_{n}$, $\omega$ is the external electric field frequency and
\begin{IEEEeqnarray}{rCl}
    R_{nm}^{abc} & \equiv & \mathrm{Im}\left[ r_{nm;a}^{b}r_{mn}^{c} \right] \ , \label{eq:RDefinition} \\
    r_{nm;a}^{b} & \equiv & \partial_{k_{a}}r_{nm}^{b} - i \left( \xi_{nn}^{a} - \xi_{mm}^{a} \right) r_{nm}^{b} \ , \label{eq:CovariantDerivative}
\end{IEEEeqnarray}
where $\xi_{nn}^{a}$ is the diagonal Berry connection. The shift current integrand can be expressed in terms of the $R$-generators using the sum rule~\cite{sipe2000second,de2020difference}
\begin{IEEEeqnarray}{rCl}
    r_{nm;b}^{a} & = & \dfrac{i}{\varepsilon_{nm}}\bigg[ \dfrac{v_{nm}^{a}\Delta_{nm}^{b}+v_{nm}^{b}\Delta_{nm}^{a}}{\varepsilon_{nm}} - w_{nm}^{ab} +  \nonumber \label{eq:SumRule} \\
    & + & \sum_{p\neq n,m}^{N}\left( \dfrac{v_{np}^{a}v_{pm}^{b}}{\varepsilon_{pm}} - \dfrac{v_{np}^{b}v_{pm}^{a}}{\varepsilon_{np}} \right) \bigg] \ ,
\end{IEEEeqnarray}
where $r_{nm}^{a} = v_{nm}^{a}/(i\varepsilon_{nm})$, $\Delta_{nm}^{a} = v_{nn}^{a}-v_{mm}^{a}$ and the connection with $R$-generators comes from
\begin{IEEEeqnarray}{rCl}
    v_{nm}^{a} & \equiv & \bra{n}\left( \partial_{k_{a}}H \right) \ket{m} = \left( \partial_{k_{a}}h_{\alpha} \right) M_{\alpha}^{nm} \ , \\
    w_{nm}^{ab} & \equiv & \bra{n} \left( \partial_{k_{a}}\partial_{k_{b}} H \right) \ket{m} = \left( \partial_{k_{a}}\partial_{k_{b}}h_{\alpha} \right) M_{\alpha}^{nm} \ .
\end{IEEEeqnarray}
The last term in the sum rule \eqref{eq:SumRule} is the one that provides the 3-generator contribution, while the rest only depend on 1- and 2-generators. The full expression of $I_{nm}^{abc}$ in terms of $R$-generators is too involved to provide further information, so we do not include it here. Nevertheless, in two-band models 3-generators do not contribute and the symmetry relation $R_{nm}^{abc}=R_{nm}^{bac}$ is accidentally satisfied. This provides another two-band constraint that relates the Berry dipole, which can be obtained from Eqs.~\eqref{eq:RDefinition} and \eqref{eq:CovariantDerivative} as
\begin{equation}
    \partial_{k_{a}}\Omega_{cb} = \sum_{\substack{n\in\mathrm{oc} \\ m\in\mathrm{unoc}}} \left( R_{nm}^{abc} - R_{nm}^{acb} \right) \ ,
\end{equation}
with the shift current integrand, and such constraint is
\begin{equation}
    S^{abc} \equiv \left( I_{12}^{abc} - I_{12}^{(abc)} \right) - \dfrac{1}{3}\left( \partial_{k_{b}}\Omega_{ca} - \partial_{k_{c}}\Omega_{ab} \right) = 0 \ , \label{eq:Identity3}
\end{equation}
where the tensor symmetrization is analogous to Eq.~\eqref{eq:SymDef}.

We now analyze the shift current coefficients of the three-band model examples. Firstly, we consider the Weyl node model with an extra band in Eq.~\eqref{eq:Ham1}. The shift current of a Weyl fermion was considered in Refs.~\cite{yang2017divergent,Osterhoudt19} with a two-band model, so that the constraint \eqref{eq:Identity3} was satisfied. By adding an extra band, we show how a more general result can be obtained. The smallest perturbative corrections in the coupling $c$ to the remote band are
\begin{equation}
    S^{xyz} = \dfrac{2c^{2}(k_{x}^{2}+k_{y}^{2})}{3k^{3}\Delta^{2}} \ .
    \label{eq:ConstraintViolation}
\end{equation}
The total shift current coefficient in this model can be computed by integrating Eq.~\eqref{eq:ShiftCurrent} and its lowest order in $\omega$ is,
\begin{equation}
    \sigma_{\mathrm{shift}}^{xyz} = \dfrac{e^{3}}{\hbar^{2}}\dfrac{c^{2}(105a^{2}+2c^{2})}{3360\pi a^{4}}\dfrac{\omega}{\Delta^{2}} \ ,
\end{equation}
similar to the behavior that is found in Ref.~\onlinecite{yang2017divergent}. However, in that work this type of term was obtained from a two-band model with quadratic corrections and hence respected the constraint \eqref{eq:Identity3}. Our result, which originates from a coupling to a remote band and explicitly breaks the constraint as seen in Eq.~\eqref{eq:ConstraintViolation}, has the same frequency scaling but can never be recovered from a two-band effective model. 

Secondly, we consider the model of a time-reversal symmetric threefold fermion in Eq.~\eqref{eq:Ham2}. The point group constraints allow only one independent shift current component that coincides with the totally symmetric one with all indices different, $\sigma_{\mathrm{shift}}^{(xyz)}$. In addition, the shift current vanishes in the linear model, and it can only receive a non-vanishing contribution from the quadratic corrections which lower the accidental symmetries of the linear model to the physical point group $T$, which are those included in Eq. \eqref{eq:Ham2}. We obtain in the small frequency limit
\begin{equation}
    \sigma_{\mathrm{shift}}^{xyz} = \dfrac{e^{3}}{\hbar^{2}}\dfrac{\sqrt{3}d}{10\pi a^{2}}\left( 1 - \dfrac{40d^{2}}{63a^{4}}\omega^{2} \right) \ .
\end{equation}
Interestingly, compared to Weyl fermions which have a shift current that vanishes linearly in $\omega$, the threefold fermion has a constant contribution at $\omega =0$.

\section{Discussion}\label{sec:Discussion}


In this work, we have introduced a method to determine any observable associated to a general finite Hamiltonian without determining its eigenstates. Our method avoids the phase arbitrariness of the eigenstates and provides analytical expressions for the minimal matrix elements required to construct any observable.



We have presented the general formalism and applied it to the case $N=3$, which is the lowest dimension that requires the use of the full machinery of SU($N$) algebra. The case of $N=4$ can also be solved exactly following the same procedure and is presented in Appendix~\ref{sec:SU(4)}. Our method becomes too impractical for larger $N$ because there is no general analytical solution for the energies and the relations between $R$-generators become more involved. However, the fact that the general case cannot be solved does not preclude that solutions might be found for concrete Hamiltonians with constraints on its SU($N$) scalars and vectors. 

Looking forward, we believe our method can be readily applied to obtain exact results in many interesting systems described by small Bloch Hamiltonians. 
One example is the structure of Berry curvature and related properties that become much richer when going form two to three bands~\cite{BBG12,LPG15}. One example where this happens is the three-band model for the CuO$_2$ planes in cuprate superconductors which features loop current ordered states~\cite{Varma97,He12}. Another example are the various three-band models featuring flat bands with non-trivial Chern numbers as those derived from the Lieb, dice or kagome lattices ~\cite{Sun11,Goldman11,Wang11}, some of which can now be engineered artificially~\cite{Slot17,Kempes19}. The $N = 4$ case is particularly interesting because often the simplest model for topological insulators (TI) require a minimum of four bands. This is the case for TIs in two~\cite{BHZ06} and three~\cite{LQZ10} dimensions, for topological crystalline insulators~\cite{Fu11} and higher-order TIs~\cite{Schindler18}.
Another field of application are synthetic systems realizing  spin-orbit coupling for spins larger than $1/2$, most notably SU(3) spin-orbit coupling~\cite{BBG12,Grass14,Hafez18}. 

Another natural area of application which we have showcased with an example is that of multifold fermions, many of which can be modeled by three or four-band Hamiltonians~\cite{Flicker18}. The transport and optical properties of systems featuring such excitations can be used to probe their topological nature~\cite{deJuan17}, and our formalism provides a natural way to model these experiments efficiently. 

A final remark is that, despite our focus on analytical results, our method will also represent an advantange in numerical calculations. For instance, the computation of observables that require loop intregrals through parameter space, like most bulk observables in crystalline solids which requires $\vec{k}$-space integrals, is generally faster if the matrix elements involved are pre-evaluated analytically. 

In summary, we have provided a method to compute the observables of small Hamliltonians via SU($N$) algebra analytically, and we believe it will find wide applications in the context of condensed matter and Bloch electrons in solids.  

\section{Acknowledgements}
We are thankful to M. A. H. Vozmediano, P. San-Jose, I. Souza and F. Pe\~naranda for support and useful conversations. We are thankful to Adolfo G. Grushin for his critical reading of the manuscript. O.P. is supported by an FPU predoctoral contract from MINECO No. FPU16/05460 and the Spanish grant PGC2018-099199-BI00 from MCIU/AEI/FEDER. F. J. acknowledges funding from the Spanish MCI/AEI through grant No. PGC2018-101988-B-C21 and from the Basque Government through grant PIBA 2019-81.

\bibliography{Manuscript}

\begin{appendices}

\section*{Appendix: Formalism}
\addcontentsline{toc}{section}{Appendix: Formalism}

\section{Summary of relations}
\addcontentsline{toc}{section}{A. Summary of relations}

This section is a collection of identities that involve the different $R$-generators up to $R=4$. These identities may be useful to obtain the analytic form of the $R$-generators or for specific physical observables that are directly related to them. Some relations are explained in the main text, but are included here for completeness.

\subsection{Relations for 1-generators}
\addcontentsline{toc}{subsection}{A.1. For 1-generators}
\label{sec:Appendix1Gen}

The diagonal matrix elements of the Hamiltonian and its powers provides relations between the 1-generators and the SU($N$) scalars. The product $\bra{n} H \ket{n}$ gives
\begin{equation}
    V_{\alpha}^{(1)}M_{\alpha}^{nn} = \varepsilon_{n} \ ,
\end{equation}
the product $\bra{n} H^{2} \ket{n}$ gives
\begin{equation}
    V_{\alpha}^{(2)}M_{\alpha}^{nn} = \varepsilon_{n}^{2} - \dfrac{2}{N}s_{2} \ ,
\end{equation}
and the product $\bra{n} H^{3} \ket{n}$ gives
\begin{equation}
    V_{\alpha}^{(3)}M_{\alpha}^{nn} = \varepsilon_{n}^{3} - \dfrac{2}{N}\left( s_{2}\varepsilon_{n} + s_{3} \right) \ .
\end{equation}

In addition, there are other 1-generator identities that can be found from the projector $P_{n}\equiv\ket{n}\bra{n}$. As it is explained in the main text, the Fierz completeness relation
\begin{equation}
    \delta_{il}\delta_{kj} = \dfrac{1}{N}\delta_{ij}\delta_{kl} + \dfrac{1}{2}\left( M_{\alpha} \right)_{ij} \left( M_{\alpha} \right)_{kl} \ ,
\end{equation}
leads to
\begin{equation}
    P_{n} = \dfrac{1}{N}\mathbb{I} + \dfrac{1}{2}M_{\alpha}^{nn}M_{\alpha} \ .
\end{equation}
The projector properties $P_{n}^{2}=P_{n}$ and $P_{n}P_{m}=0$ lead to
\begin{IEEEeqnarray}{rCl}
    M_{\alpha}^{nn}M_{\alpha}^{mm} & = & 2\delta_{nm} - \dfrac{2}{N} , \\
    d_{\alpha\beta\gamma}M_{\beta}^{nn}M_{\gamma}^{mm} & = & 2\delta_{nm}M_{\alpha}^{nn} - \dfrac{2}{N}\left( M_{\alpha}^{nn} + M_{\alpha}^{mm} \right) , \qquad
\end{IEEEeqnarray}
respectively.

\subsection{Relations for 2-generators}
\addcontentsline{toc}{subsection}{A.2. For 2-generators}

The diagonal matrix elements of the product of two SU($N$) generators, $\bra{n}M_{\alpha}M_{\beta}\ket{n}$, gives a relation between 1- and 2-generators if we introduce a resolution of the identity
\begin{IEEEeqnarray}{rCl}
    \sum_{m=1}^{N} M_{\alpha}^{nm}M_{\beta}^{mn} & = & \dfrac{2}{N}\delta_{\alpha\beta} + S_{\alpha\beta\gamma}M_{\gamma}^{nn} \ .
    \label{eq:Appendix-Easiest-2gen-1gen}
\end{IEEEeqnarray}
Alternatively, we can consider larger products of SU($N$) generators and contract the remaining indices with the $V_{\alpha}^{(r)}$ vectors. For instance, the product $V_{\gamma}^{(r)}\bra{n}M_{\alpha}M_{\gamma}M_{\beta}\ket{n}$ gives
\begin{IEEEeqnarray}{rCl}
    \sum_{m=1}^{N}\left( V_{\gamma}^{(r)}M_{\gamma}^{mm} \right) M_{\alpha}^{nm}M_{\beta}^{mn} & = & \dfrac{2}{N} \left( V_{\alpha}^{(r)}M_{\beta}^{nn} + S_{\alpha\gamma\beta}V_{\gamma}^{(r)} \right) \nonumber \\
    & + & S_{\alpha\gamma\sigma}S_{\sigma\beta\rho}V_{\gamma}^{(r)}M_{\rho}^{nn} \ ,
    \label{eq:Appendix-ReMM-E}
\end{IEEEeqnarray}
where the left-hand side can be evaluated from the 1-generator identities of the previous subsection.

The non-diagonal matrix elements $\bra{n} \big[ M_{\alpha} , H \big] \ket{m}$ and $\bra{n}\left\{ M_{\alpha} , H \right\} \ket{m}$ give
\begin{IEEEeqnarray}{rCl}
    M_{\alpha}^{nm} & = & \dfrac{2if_{\alpha\sigma\rho}h_{\sigma}}{\varepsilon_{m}-\varepsilon_{n}}M_{\rho}^{nm} \label{eq:GaugeDep1} \ , \\
    M_{\alpha}^{nm} & = & \dfrac{2d_{\alpha\sigma\rho}h_{\sigma}}{\varepsilon_{n}+\varepsilon_{m}}M_{\rho}^{nm} \ , \label{eq:GaugeDep2}
\end{IEEEeqnarray}
respectively. These gauge-dependent identities multiplied by $M_{\beta}^{mn}$ relate the real and imaginary parts of 2-generators
\begin{IEEEeqnarray}{rCl}
    \mathrm{Re}\big[ M_{\alpha}^{nm}M_{\beta}^{mn} \big] & = & \dfrac{2d_{\alpha\sigma\rho}h_{\sigma}}{\varepsilon_{n}+\varepsilon_{m}}\mathrm{Re}\big[ M_{\rho}^{nm}M_{\beta}^{mn} \big] \ , \quad \\
    \mathrm{Re}\big[ M_{\alpha}^{nm}M_{\beta}^{mn} \big] & = & \dfrac{2f_{\alpha\sigma\rho}h_{\sigma}}{\varepsilon_{n}-\varepsilon_{m}}\mathrm{Im}\big[ M_{\rho}^{nm}M_{\beta}^{mn} \big] \ , \quad \\
    \mathrm{Im}\big[ M_{\alpha}^{nm}M_{\beta}^{mn} \big] & = & \dfrac{2d_{\alpha\sigma\rho}h_{\sigma}}{\varepsilon_{n}+\varepsilon_{m}}\mathrm{Im}\big[ M_{\rho}^{nm}M_{\beta}^{mn} \big] \ , \quad \\
    \mathrm{Im}\big[ M_{\alpha}^{nm}M_{\beta}^{mn} \big] & = & \dfrac{2f_{\alpha\sigma\rho}h_{\sigma}}{\varepsilon_{m}-\varepsilon_{n}}\mathrm{Re}\big[ M_{\rho}^{nm}M_{\beta}^{mn} \big] \ , \quad 
    \label{eq:Appendix-ReMM-to-ImMM}
\end{IEEEeqnarray}
which can be easily generalized to higher order $R$-generators. Higher powers of the Hamiltonian can be used to obtain different relations involving higher powers of the energy $\varepsilon_{n}$.

\subsection{Relations for 3-generators}
\addcontentsline{toc}{subsection}{A.3. For 3-generators}

The diagonal matrix elements of the product of three SU($N$) generators, $\bra{n}M_{\alpha}M_{\beta}M_{\gamma}\ket{n}$, gives a relation between 2- and 3-generators if we introduce a resolution of the identity
\begin{IEEEeqnarray}{rCl}
    \sum_{m,p=1}^{N} M_{\alpha}^{nm}M_{\beta}^{mp}M_{\gamma}^{pn} & = & \dfrac{2}{N}\delta_{\alpha\beta}M_{\gamma}^{nn} + \nonumber \\
    & + & S_{\alpha\beta\sigma}\sum_{m=1}^{N}M_{\sigma}^{nm}M_{\gamma}^{mn} .
\end{IEEEeqnarray}
Alternatively, we can consider larger products of SU($N$) generators and contract the remaining indices with $V_{\alpha}^{(r)}$ vectors. For instance, the product $V_{\lambda}^{(r)}\bra{n} M_{\alpha}M_{\lambda}M_{\beta}M_{\gamma} \ket{n}$ gives
\begin{IEEEeqnarray}{rCl}
    \sum_{m,p=1}^{N}\left( V_{\lambda}^{(r)}M_{\lambda}^{mm} \right) & & M_{\alpha}^{nm}M_{\beta}^{mp}M_{\gamma}^{pn} = \dfrac{2}{N}S_{\alpha\lambda\beta}V_{\lambda}^{(r)}M_{\gamma}^{nn} \nonumber \\
    & & + \dfrac{2}{N}V_{\alpha}^{(r)}\sum_{m=1}^{N}M_{\beta}^{nm}M_{\gamma}^{mn} \nonumber \\
    & & + S_{\alpha\lambda\tau}S_{\tau\beta\sigma}V_{\lambda}^{(r)}\sum_{m=1}^{N}M_{\sigma}^{nm}M_{\gamma}^{mn} . \quad
\end{IEEEeqnarray}

These last examples illustrate how this kind of relations become too impractical for higher order $R$-generators. As it is discussed in the main text, the non-diagonal matrix elements are the simplest way to derive relations between 2- and 3-generators. For instance, considering $n\neq m$, the product $\bra{n}M_{\alpha}M_{\beta}\ket{m}$ multiplied by $M_{\gamma}^{mn}$ leads to
\begin{IEEEeqnarray}{rCl}
    \sum_{l=1}^{N} M_{\alpha}^{nl}M_{\beta}^{lm}M_{\gamma}^{mn} & = & S_{\alpha\beta\sigma} M_{\sigma}^{nm}M_{\gamma}^{mn} \ .
    \label{eq:Appendix-Easiest-3gen-2gen}
\end{IEEEeqnarray}
We can also consider larger gauge-dependent products of SU($N$) generators and contract the remaining indices with the $V_{\alpha}^{(r)}$ vectors. For instance, the product $V_{\lambda}^{(r)}\bra{n}M_{\alpha}M_{\lambda}M_{\beta}\ket{m}$ multiplied by $M_{\gamma}^{mn}$ gives
\begin{IEEEeqnarray}{rCl}
    \sum_{p=1}^{N} \left( V_{\lambda}^{(r)}M_{\lambda}^{pp} \right) M_{\alpha}^{np} & & M_{\beta}^{pm}M_{\gamma}^{mn} = \dfrac{2}{N}V_{\alpha}^{(r)}M_{\beta}^{nm}M_{\gamma}^{mn} + \nonumber \\
    & & + S_{\alpha\lambda\sigma}S_{\sigma\beta\rho}V_{\lambda}^{(r)}M_{\rho}^{nm}M_{\gamma}^{mn} . \label{eq:3-to-2-gen-E}
\end{IEEEeqnarray}

Finally, the real and imaginary parts of 3-generators are related through Eqs.~\eqref{eq:GaugeDep1} and \eqref{eq:GaugeDep2} when they are multiplied by $M_{\beta}^{ml}M_{\gamma}^{ln}$. The relations are
\begin{IEEEeqnarray}{rCl}
    \mathrm{Re}\big[ M_{\alpha}^{nm}M_{\beta}^{ml}M_{\gamma}^{ln} \big] & = & \dfrac{2d_{\alpha\sigma\rho}h_{\sigma}}{\varepsilon_{n}+\varepsilon_{m}}\mathrm{Re}\big[ M_{\rho}^{nm}M_{\beta}^{ml}M_{\gamma}^{ln} \big] , \qquad \\
    \mathrm{Re}\big[ M_{\alpha}^{nm}M_{\beta}^{ml}M_{\gamma}^{ln} \big] & = & \dfrac{2f_{\alpha\sigma\rho}h_{\sigma}}{\varepsilon_{n}-\varepsilon_{m}}\mathrm{Im}\big[ M_{\rho}^{nm}M_{\beta}^{ml}M_{\gamma}^{ln} \big] , \qquad \\
    \mathrm{Im}\big[ M_{\alpha}^{nm}M_{\beta}^{ml}M_{\gamma}^{ln} \big] & = & \dfrac{2d_{\alpha\sigma\rho}h_{\sigma}}{\varepsilon_{n}+\varepsilon_{m}}\mathrm{Im}\big[ M_{\rho}^{nm}M_{\beta}^{ml}M_{\gamma}^{ln} \big] , \qquad \\
    \mathrm{Im}\big[ M_{\alpha}^{nm}M_{\beta}^{ml}M_{\gamma}^{ln} \big] & = & \dfrac{2f_{\alpha\sigma\rho}h_{\sigma}}{\varepsilon_{m}-\varepsilon_{n}}\mathrm{Re}\big[ M_{\rho}^{nm}M_{\beta}^{ml}M_{\gamma}^{ln} \big] . \qquad 
    \label{eq:Appendix-ReMM-to-ImMM}
\end{IEEEeqnarray}

\subsection{Relations for 4-generators}
\addcontentsline{toc}{subsection}{A.4. For 4-generators}

The diagonal matrix elements of the product of four SU($N$) generators, $\bra{n}M_{\alpha}M_{\beta}M_{\gamma}M_{\delta}\ket{n}$, gives a relation between 3- and 4-generators if we introduce a resolution of the identity
\begin{IEEEeqnarray}{rCl}
    \sum_{l,m,p=1}^{N} M_{\alpha}^{nl}M_{\beta}^{lm}M_{\gamma}^{mp} & & M_{\delta}^{pn} = \dfrac{2}{N}\delta_{\alpha\beta}\sum_{m=1}^{N}M_{\gamma}^{nm}M_{\delta}^{mn}+\qquad \nonumber \\
    & & + S_{\alpha\beta\sigma}\sum_{m,p=1}^{N}M_{\sigma}^{nm}M_{\gamma}^{mp}M_{\delta}^{pn} \ .
\end{IEEEeqnarray}
Alternatively, we can consider larger products of SU($N$) generators and contract the remaining indices with $V_{\alpha}^{(r)}$ vectors. For instance, the product $V_{\lambda}^{(r)}\bra{n} M_{\alpha}M_{\lambda}M_{\beta}M_{\gamma}M_{\delta}\ket{n}$ gives
\begin{IEEEeqnarray}{rCl}
    \sum_{l,m,p=1}^{N} \big( V_{\lambda}^{(r)} & & M_{\lambda}^{ll} \big) M_{\alpha}^{nl}M_{\beta}^{lm}M_{\gamma}^{mp}M_{\delta}^{pn} = \nonumber \\
    & & \dfrac{2}{N}S_{\alpha\lambda\beta}V_{\lambda}^{(r)}\sum_{m=1}^{N}M_{\gamma}^{nm}M_{\delta}^{mn}+ \nonumber \\
    & & + \dfrac{2}{N}V_{\alpha}^{(r)}\sum_{m,p=1}^{N}M_{\beta}^{nm}M_{\gamma}^{mp}M_{\delta}^{pn}+ \nonumber \\
    & & + S_{\alpha\lambda\sigma}S_{\sigma\beta\rho}\sum_{m,p=1}^{N}M_{\rho}^{nm}M_{\gamma}^{mp}M_{\delta}^{pn} \ .
\end{IEEEeqnarray}

These last examples show again how this kind of relations become too impractical for higher order $R$-generators. The non-diagonal matrix elements are again the simplest way to derive relations between 3- and 4-generators. For instance, considering $n\neq m$, the product $\bra{n} M_{\alpha}M_{\beta} \ket{m}$ multiplied by $M_{\gamma}^{mp}M_{\delta}^{pn}$ with $p\neq n,m$ leads to
\begin{IEEEeqnarray}{rCl}
    \sum_{l=1}^{N} M_{\alpha}^{nl}M_{\beta}^{lm}M_{\gamma}^{mp}M_{\delta}^{pn} & = & S_{\alpha\beta\sigma}M_{\sigma}^{nm}M_{\gamma}^{mp}M_{\delta}^{pn} \ . \quad
    \label{eq:4-to-3-gen}
\end{IEEEeqnarray}
We can also consider larger gauge-dependent products of SU($N$) generators and contract the remaining indices with the $V_{\alpha}^{(r)}$ vectors. For instance, the product $V_{\lambda}^{(r)}\bra{n} M_{\alpha}M_{\lambda}M_{\beta} \ket{m}$ multiplied by $M_{\gamma}^{ml}M_{\delta}^{ln}$ with $l\neq n,m$ gives
\begin{IEEEeqnarray}{rCl}
    \sum_{p=1}^{N} \left( V_{\lambda}^{(r)}M_{\lambda}^{pp} \right) & & M_{\alpha}^{np} M_{\beta}^{pm}M_{\gamma}^{ml}M_{\delta}^{ln} = \nonumber \\
    & & \dfrac{2}{N}V_{\alpha}^{(r)}M_{\beta}^{nm}M_{\gamma}^{ml}M_{\delta}^{ln} + \nonumber \\
    & & + S_{\alpha\lambda\sigma}S_{\sigma\beta\rho}V_{\lambda}^{(r)}M_{\rho}^{nm}M_{\gamma}^{ml}M_{\delta}^{ln} . \quad \label{eq:4-to-3-gen-E}
\end{IEEEeqnarray}

Finally, the real and imaginary parts of 4-generators are related through Eqs.~\eqref{eq:GaugeDep1} and \eqref{eq:GaugeDep2} as in the 3-generator case explained in the last subsection, but multiplying them by $M_{\beta}^{ml}M_{\gamma}^{lp}M_{\delta}^{pn}$ instead of $M_{\beta}^{ml}M_{\gamma}^{ln}$. We do not include them here since they do not provide further information.

\section{ORTHOGONALITY RELATIONS}
\addcontentsline{toc}{subsection}{B. ORTHOGONALITY RELATIONS}

We identify two kinds of orthogonality relations for $R$-generators. The first one comes from non-diagonal matrix elements as $\bra{n} H \ket{m} = 0$, which directly implies that
\begin{equation}
    h_{\alpha}M_{\alpha}^{nm} = 0 \ .
\end{equation}
Although this relation is gauge-dependent, it can be multiplied by any string of matrix elements that compensates such gauge dependence, revealing an orthogonality relation between $h_{\alpha}$ and $R$-generators with $R\geq2$. Moreover, this orthogonality extends to any other vector $V_{\alpha}^{(r)}$, defined in Eq.~\eqref{eq:SUN_Vectors}, considering the non-diagonal matrix element $\bra{n} H^{r} \ket{m}$, which leads to
\begin{equation}
    V_{\alpha}^{(r)}M_{\alpha}^{nm} = 0 \ .
\end{equation}
Therefore we conclude that the $R$-generators with $R\geq2$ are orthogonal to any vector $V_{\alpha}^{(r)}$.

The second kind of orthogonality relations are specially useful in Bloch Hamiltonians, since they involve the momentum derivative of 1-generators. The simplest example can be checked by explicit evaluation
\begin{equation}
    h_{\alpha}\left( \partial_{k_{a}}M_{\alpha}^{nn} \right) = \varepsilon_{n}\partial_{k_{a}}\left( \braket{n|n} \right) = 0 \ .
\end{equation}
This relation easily extends to the rest of $V_{\alpha}^{(r)}$ vectors, replacing the scalar $\varepsilon_{n}$ by the one that results from evaluating $V_{\alpha}^{(r)}M_{\alpha}^{nn}$. Therefore we conclude that 1-generators obey the following set of orthogonality relations
\begin{equation}
    V_{\alpha}^{(r)}\left( \partial_{k_{a}}M_{\alpha}^{nn} \right) = 0 \ .
\end{equation}
Note that this reasoning does not apply exclusively to momentum derivatives but to derivatives of any other Hamiltonian parameter.

\section{SU(4) SOLUTION}
\addcontentsline{toc}{subsection}{C. SU(4) SOLUTION}
\label{sec:SU(4)}
Here we apply the general strategy described in Section~\ref{sec:GeneralStrategy} to the case $N=4$. The starting point is the closing $d$-family identity for $N=4$, which is~\cite{de1998invariant}
\begin{equation}
    d^{(5)}_{(\alpha_{1}\alpha_{2}\alpha_{3}\alpha_{4}\alpha_{5})} = \dfrac{2}{3}d_{(\alpha_{1}\alpha_{2}\alpha_{3}}\delta_{\alpha_{4}\alpha_{5})}  \ ,
    \label{eq:d-family-SU4}
\end{equation}
where the tensor symmetrization is defined in Eq.~\eqref{eq:SymDef}. The characteristic polynomial \eqref{eq:CharacPol} for $N=4$,
\begin{equation}
    p_{H}(4,\varepsilon_{n}) = \varepsilon_{n}^{4} - s_{2}\varepsilon_{n}^{2} - \dfrac{2}{3}s_{3} \varepsilon_{n} + \dfrac{1}{4}\left( s_{2}^{2} - 2s_{4} \right) , \quad \label{eq:ChaEqSU4}
\end{equation}
has an intricate but analytic solution
\begin{IEEEeqnarray}{rCl}
    \varepsilon_{1,2} & = & -S \pm \dfrac{1}{2}\sqrt{-4S^{2} + 2s_{2} - \dfrac{2s_{3}}{3S}}  \  ,  \\
    \varepsilon_{3,4} & = & S \pm \dfrac{1}{2}\sqrt{-4S^{2} + 2s_{2} + \dfrac{2s_{3}}{3S}}  \  , 
\end{IEEEeqnarray}
given in terms of the parameters
\begin{IEEEeqnarray}{rCl}
    S & \equiv & \dfrac{1}{2}\sqrt{\dfrac{2s_{2}}{3} + \dfrac{2}{3}\Delta_{0}^{1/2}\cos\left( \dfrac{\phi}{3} \right)}  \ ,  \\
    \phi & \equiv & \arccos\left( \dfrac{\Delta_{1}}{2\Delta_{0}^{3/2}} \right) \  , \\
    \Delta_{0} & \equiv & 2\left( 2 s_{2}^{2} - 3s_{4} \right)  \  ,  \\
    \Delta_{1} & \equiv & 4\left( 4 s_{2}^{3} + 3 s_{3}^{2} - 9s_{2}s_{4} \right)  \  .
\end{IEEEeqnarray}

To determine the 1-generators, there are only three vectors available, as it is indicated in Eq.~\eqref{eq:SUN_Vectors}. We need the specific tensor contraction
\begin{IEEEeqnarray}{rCl}
    d_{\alpha\beta\gamma}h_{\beta}V_{\gamma}^{(3)} & = & \dfrac{1}{6}s_{3}V_{\alpha}^{(1)} + \dfrac{1}{2}s_{2}V_{\alpha}^{(2)} \ ,
\end{IEEEeqnarray}
which comes from the closing $d$-family identity \eqref{eq:d-family-SU4} and the general SU($N$) identities \eqref{eq:fdCyclic}-\eqref{eq:ff-dd}. Then the vector decomposition of $M_{\alpha}^{nn}$ in Eq.~\eqref{eq:MainEq} leads to the following expression for the 1-generators
\begin{equation}
    M_{\alpha}^{nn} = 3\dfrac{\left( \varepsilon_{n}^{2}-s_{2}/2 \right) V_{\alpha}^{(1)} + \varepsilon_{n}V_{\alpha}^{(2)} + V_{\alpha}^{(3)}}{ 6\varepsilon_{n}\left( \varepsilon_{n}^{2}-s_{2}/2 \right) - s_{3}} .
\end{equation}

The mathematical structure becomes more intricate for higher order $R$-generators, but the strategy results as simple as for SU(3) using suitable tensor definitions. For the 2-generators we define
\begin{IEEEeqnarray}{rCl}
    A_{\alpha\beta}^{(n)} & \equiv & \dfrac{1}{2}\delta_{\alpha\beta} + d_{\alpha\beta\gamma}M_{\gamma}^{nn} - M_{\alpha}^{nn}M_{\beta}^{nn} , \\
    B_{\alpha\beta}^{(n)} & \equiv & \dfrac{1}{2}\left( h_{\alpha}M_{\beta}^{nn} + d_{\alpha\beta\gamma}h_{\gamma} \right) - \varepsilon_{n}M_{\alpha}^{nn}M_{\beta}^{nn} + \nonumber \\ & & + \left( d_{\alpha\rho\gamma}d_{\beta\rho\delta} - f_{\alpha\rho\gamma}f_{\beta\rho\delta} \right) h_{\gamma} M_{\delta}^{nn} , \\
    C_{\alpha\beta}^{(n)} & \equiv & \dfrac{s_{2}}{2}A_{\alpha\beta}^{(n)} + \dfrac{1}{2}\left( V_{\alpha}^{(2)}M_{\beta}^{nn}+d_{\alpha\beta\gamma}V_{\gamma}^{(2)} \right) + \nonumber \\
    & & + \left( d_{\alpha\rho\gamma}d_{\beta\rho\delta}-f_{\alpha\rho\gamma}f_{\beta\rho\delta} \right) V_{\gamma}^{(2)}M_{\delta}^{nn} - \nonumber \\ 
    & & - \left( \varepsilon_{n}^{2}-s_{2}/2 \right) M_{\alpha}^{nn}M_{\beta}^{nn} . \qquad
\end{IEEEeqnarray}
Then the real part of Eqs.~\eqref{eq:Appendix-Easiest-2gen-1gen} and\eqref{eq:Appendix-ReMM-E} with $r=1,2$ simplify to the following system of equations
\begin{IEEEeqnarray}{rCl}
    \sum_{m\neq n}^{N}\mathrm{Re}\left[ M_{\alpha}^{nm}M_{\beta}^{mn} \right] ) & = & A_{\alpha\beta}^{(n)} \ , \\
    \sum_{m\neq n}^{N} \varepsilon_{n} \mathrm{Re}\left[ M_{\alpha}^{nm}M_{\beta}^{mn} \right] & = & B_{\alpha\beta}^{(n)} \ , \\
    \sum_{m\neq n}^{N}\left( \varepsilon_{n}^{2}-\dfrac{s_{2}}{2} \right) \mathrm{Re}\left[ M_{\alpha}^{nm}M_{\beta}^{mn} \right] & = & C_{\alpha\beta}^{(n)} - \dfrac{s_{2}}{2}A_{\alpha\beta}^{(n)} \ . \qquad
\end{IEEEeqnarray}
This set leads to the following analytic expression for the real part of 2-generators
\begin{equation}
    \mathrm{Re}\left[ M_{\alpha}^{nm}M_{\beta}^{mn} \right] = \dfrac{\varepsilon_{l}\varepsilon_{p}A_{\alpha\beta}^{(n)} + \left( \varepsilon_{n}+\varepsilon_{m} \right) B_{\alpha\beta}^{(n)} + C_{\alpha\beta}^{(n)}}{\left( \varepsilon_{m}-\varepsilon_{l} \right) \left( \varepsilon_{m}-\varepsilon_{p} \right)} , 
    \label{eq:SU(4)-2gen}
\end{equation}
where $(n,m,l,p)$ must be all different indices. The imaginary part of the 2-generators is obtained in terms of the real one using Eq.~\eqref{eq:Appendix-ReMM-to-ImMM}.

For the real part of the 3-generators, we define the following tensors
\begin{widetext}
\begin{IEEEeqnarray}{rCl}
    T_{\alpha\beta\gamma}^{(nm)} & \equiv & d_{\alpha\beta\sigma}\mathrm{Re}\left[ M_{\sigma}^{nm}M_{\gamma}^{mn} \right] - f_{\alpha\beta\sigma}\mathrm{Im}\left[ M_{\sigma}^{nm}M_{\gamma}^{mn} \right] - M_{\alpha}^{nn}\mathrm{Re}\left[ M_{\beta}^{nm}M_{\gamma}^{mn} \right] - M_{\beta}^{mm}\mathrm{Re}\left[ M_{\alpha}^{nm}M_{\gamma}^{mn} \right]  \  ,  \\
    Q_{\alpha\beta\gamma}^{(nm)} & \equiv & \left( d_{\alpha\sigma\delta}d_{\delta\beta\rho} - f_{\alpha\sigma\delta}f_{\delta\beta\rho} \right) h_{\sigma}\mathrm{Re}\left[ M_{\rho}^{nm}M_{\gamma}^{mn} \right] - \left( d_{\alpha\sigma\delta}f_{\delta\beta\rho} + f_{\alpha\sigma\delta}d_{\delta\beta\rho} \right) h_{\sigma}\mathrm{Im}\left[ M_{\rho}^{nm}M_{\gamma}^{mn} \right] - \nonumber \\
    & & - \left( \varepsilon_{n}M_{\alpha}^{nn}-h_{\alpha}/2 \right) \mathrm{Re}\left[ M_{\beta}^{nm}M_{\gamma}^{mn} \right] - \varepsilon_{m}M_{\beta}^{mm}\mathrm{Re}\left[ M_{\alpha}^{nm}M_{\gamma}^{mn} \right] \ ,
\end{IEEEeqnarray}
\end{widetext}
so we can rewrite Eqs.~\eqref{eq:Appendix-Easiest-3gen-2gen} and \eqref{eq:3-to-2-gen-E} with $r=1$ as
\begin{IEEEeqnarray}{rCl}
    \sum_{l\neq n,m}^{4} \mathrm{Re}\left[ M_{\alpha}^{nl}M_{\beta}^{lm}M_{\gamma}^{mn} \right] & = & T_{\alpha\beta\gamma}^{(nm)}  \  , \\
    \sum_{l\neq n\neq m}^{4} \varepsilon_{l} \ \mathrm{Re}\left[ M_{\alpha}^{nl}M_{\beta}^{lm}M_{\gamma}^{mn} \right] & = & Q_{\alpha\beta\gamma}^{(nm)}  \  ,
\end{IEEEeqnarray}
where $n\neq m$, and get the following analytic expression for the real part of the 3-generators
\begin{IEEEeqnarray}{rCl}
    \mathrm{Re}\left[ M_{\alpha}^{nl}M_{\beta}^{lm}M_{\gamma}^{mn} \right] & = & \dfrac{Q_{\alpha\beta\gamma}^{(nm)} + \left( \varepsilon_{l}+\varepsilon_{n}+\varepsilon_{m} \right) T_{\alpha\beta\gamma}^{(nm)}}{2\varepsilon_{l}+\varepsilon_{n}+\varepsilon_{m}} . \qquad
\end{IEEEeqnarray}
The imaginary part of the 3-generators is obtained in terms of the real one using Eq.~\eqref{eq:ReMMM-to-ImMMM}.

Finally, the real and imaginary parts of the 4-generators can be obtained from the real part of Eq.~\eqref{eq:4-to-3-gen} and from \eqref{eq:GaugeDep1}, which lead to the following relations where $(n,m,l,p)$ are all different indices
\begin{widetext}
\begin{IEEEeqnarray}{rCl}
    \mathrm{Re}\left[ M_{\alpha}^{nl}M_{\beta}^{lm}M_{\gamma}^{mp}M_{\delta}^{pn} \right] & = &  d_{\alpha\beta\sigma}\mathrm{Re}\left[ M_{\sigma}^{nm}M_{\gamma}^{mp}M_{\delta}^{pn}\right] - f_{\alpha\beta\sigma}\mathrm{Im}\left[ M_{\sigma}^{nm}M_{\gamma}^{mp}M_{\delta}^{pn}\right] - \nonumber \\
    & & - M_{\alpha}^{nn}\mathrm{Re}\left[ M_{\beta}^{nm}M_{\gamma}^{mp}M_{\delta}^{pn} \right] - M_{\beta}^{mm}\mathrm{Re}\left[ M_{\alpha}^{nm}M_{\gamma}^{mp}M_{\delta}^{pn} \right] - \nonumber \\
    & & - \mathrm{Re}\big[ M_{\alpha}^{np}M_{\delta}^{pn} \big] \ \mathrm{Re}\big[ M_{\gamma}^{mp}M_{\beta}^{pm} \big] + \mathrm{Im}\big[ M_{\alpha}^{np}M_{\delta}^{pn} \big] \ \mathrm{Im}\big[ M_{\gamma}^{mp}M_{\beta}^{pm} \big]  , \quad \\
    \mathrm{Im}\big[ M_{\alpha}^{nm}M_{\beta}^{ml}M_{\gamma}^{lp}M_{\delta}^{pn} \big] & = & \dfrac{2if_{\alpha\sigma\rho}h_{\sigma}}{\varepsilon_{m}-\varepsilon_{n}} \mathrm{Re}\big[ M_{\rho}^{nm}M_{\beta}^{ml}M_{\gamma}^{lp}M_{\delta}^{pn} \big] . \quad
    \label{eq:Appendix-ReMMMM-to-ImMMMM}
\end{IEEEeqnarray}
\end{widetext}

\end{appendices}

\end{document}